# Janus: A Systems Engineering Approach to the Design of Industrial Cyber-Physical Systems

Dennis Jarvis[1], Jacqueline Jarvis[2], Chen-Wei Yang[3], Roopak Sinha[4],
& Valeriy Vyatkin[5]

[1&2]School of Engineering and Technology, Central Queensland University, Brisbane, Australia
[3]Department of Computer Science, Electrical and Space Engineering,
Luleå University of Technology, Luleå, Sweden
[4]IT & Software Engineering, Auckland University of Technology, Auckland, New Zealand
[5]Department of Computer Science, Electrical and Space Engineering,
Luleå University of Technology, Luleå, Sweden
[5]Department of Electrical Engineering and Automation, Aalto University, Espoo, Finland
emails: [1]d.jarvis@cqu.edu.au, [2]j.jarvis@cqu.edu.au, [3]chen-wei.yang@ltu.se, [4]roopak.sinha@aut.ac.nz,
[5]vyatkin@ieee.org

**Abstract**

*The benefits that arise from the adoption of a systems engineering approach to the design of engineered systems are well understood and documented. However, with software systems, different approaches are required given the changeability of requirements and the malleability of software. With the design of industrial cyber-physical systems, one is confronted with the challenge of designing engineered systems that have a significant software component. Furthermore, that software component must be able to seamlessly interact with both the enterprise's business systems and industrial systems. In this paper, we present Janus, which together with the GORITE BDI agent framework, provides a methodology for the design of agent-based industrial cyber-physical systems. Central to the Janus approach is the development of a logical architecture as in traditional systems engineering and then the allocation of the logical requirements to a BDI (Belief Desire Intention) agent architecture which is derived from the physical architecture for the system. Janus has its origins in product manufacturing; in this paper, we apply it to the problem of Fault Location, Isolation and Service Restoration (FLISR) for power substations.*

**Keywords:** cyber-physical systems, multi-agent systems, industrial automation, IEC 61850, IEC 61499

## 1. INTRODUCTION

Our interest lies in the development of multi-agent architectures and frameworks that can be deployed at all levels of an industrial enterprise. However as noted in [1], the lack of design methodologies and software frameworks that directly support the development of agent-based applications is limiting the uptake of agent technology in the industry.

Central to the operations of industrial enterprises are the physical systems that produce the product or generate the service that the enterprise provides. These physical systems need to interact with the supporting business systems, such as order management, where changes in customer orders impact on production and production disruptions can impact on customer orders. Agent technology provides an attractive means to deliver the required integration, but more importantly, to provide the flexibility to handle change both at the business level and at the production level. However, how should agent-enabled systems be designed? In answer to this question, we observe that agent-based production systems are amenable to design using standard Systems Engineering practices if:

1. Functional decomposition and analysis is reformulated as goal decomposition and analysis
2. Agents and teams of agents are mapped to the systems, subsystems and leaf nodes of the physical architecture
3. Functional allocation is reformulated as goal allocation to the agent/teams of agents identified in 2.

The resulting design methodology is called Janus and is described in Section 3. However, its efficacy will be greatly improved if the goal allocation process can occur in such a way that executable agent behaviours are generated, rather than paper-based specifications as in traditional Systems Engineering. In this regard, Janus is intended to be used in conjunction with the GORITE BDI framework [2]. Janus draws from various sources and case studies in the domain of agent-based manufacturing; this background material forms Section 2 and Janus is presented in Section 3. In section 4, the applicability of Janus to a quite different cyber-physical application, namely power substation automation, is examined; the case study is then discussed in Section 5. Concluding remarks are provided in Section 6.

## 2. Background

Central to Janus is the alignment of the logical (goal) architecture with the agent architecture and the alignment of the agent architecture with the physical architecture. The latter mapping implies an agent architecture in which teams of agents are explicitly represented and correspond to systems and/or subsystems. Individual agents in turn map to subsystem components. For example, in [3], the cell is the team/system and its constituent machines are the agents/components. Such a modelling perspective fosters reusability, as each entity, be it a system, subsystem or component is an entity in its own right as well as potentially being a component in a larger system. This duality of the system/sub-system relationship was first documented by Koestler [4] within the context of biological and social systems. Koestler used the so-called Janus effect as a metaphor for describing this dichotomy:

"*like the Roman god Janus, members of a hierarchy have two faces looking in opposite directions*"

In other words, these members can be thought of as self-contained wholes looking downwards to the subordinate level and/or as dependent parts looking upward. Koestler coined the terms holon to refer to system/subsystem and holarchy to describe the organisational structure formed by holons. His work subsequently provided the conceptual basis for the Holonic Manufacturing Systems (HMS) project [5]. However, as was noted in [6], the HMS Project focused on the concept of a holon as a distinct cyber-physical entity and not on the concept of holarchy.

In keeping with the prevailing view of the multi-agent systems community, holarchies were viewed as dynamic constructs formed in response to a need to achieve particular goals. However, the dynamism extended only to holarchy formation (determining the structure of a team of holons that could achieve a particular goal) and not to actual team plan formation. While in many situations, dynamic planning is essential, the success of the Belief-Desire-Intention (BDI) model of agency [7] has demonstrated that in many applications it is not. Rather, it is often the case that providing an agent with the ability to choose between different predefined courses of action to achieve a particular goal is sufficient. Nonetheless, the HMS project did not embrace the BDI model, preferring to leave plan definition and execution as implementation concerns. Furthermore, while the achievement of complex goals often requires dynamic team formation, such team formation generally occurs within an existing organisational structure, such as a factory which has been organised into manufacturing cells Neither the HMS Project nor the BDI frameworks of the time addressed this issue and the top-down approach to team formation that it engenders.

Since the completion of the HMS Project, the concept of holarchy has been captured in the BDI agent frameworks JACK Teams [8] and GORITE [2], developed by Rönnquist. In these frameworks,

4. a holarchy is mapped to a hierarchical structuring of teams, sub-teams and individual agents

5. as with an individual agent, an agent team has its own beliefs, desires and intentions

6. team behaviour is specified in terms of roles (defined as collections of related goals)

7. team behaviour is realised by selecting team members that are able to perform the roles required by the behaviour.

Both JACK Teams and GORITE have been successfully employed in the development of agent-based execution systems for manufacturing cells [3], [9], [10]. In all cases, teams of agents were employed, with individual agents mapped to PLC controlled machines/devices (IEC 61131 in [3]; IEC 61499 in [9], [10]). Additionally, a holarchic design using JACK Teams was developed for the Cambridge Holonic Enterprise Demonstrator [11] but was ultimately implemented using JACK [8] because of schedule pressures [12].

While the BDI systems implemented above embodied the holarchic concept espoused by Koestler, they were developed in the absence of any explicit design methodology. In this regard, note that while the Prometheus methodology [13] and its associated design tool [14] haves been widely used in the development of BDI agent applications, Prometheus does not support agent teams. Furthermore, while the design tool has code generation capabilities, code is restricted to the proprietary JACK framework. Consequently, it was decided to formulate a systematic design methodology called Janus which is aligned with the open source GORITE BDI framework. The formulation of Janus has been informed by reflection on the experiences gained through the development of execution systems developed using JACK, JACK Teams and GORITE and from the HMS industry testbeds, in particular [15].

In traditional systems engineering [16], three design phases are enunciated – Conceptual Design, Preliminary Design and Detailed Design and Development. In the first of these phases, business needs and system requirements are identified and a logical architecture is constructed through a process of functional decomposition and analysis. In the second phase a candidate physical architecture is identified at the subsystem level and requirements embodied in the logical architecture are refined and then allocated to subsystems, taking into account the technical performance measures associated with each requirement. Further refinement of both the logical and physical architectures to the component level then occurs in the final design phase, resulting in a fully specified physical architecture. Note that while design proceeds in a top-down manner, it will often be the case that the physical architecture will be constrained by the nature of the system, as with factories that employ manufacturing cells and other machine aggregations. This is also the case with power substations, as will be seen in Section 4. Also, as the design process progresses, there will be a need for specialist engineering and design input.

These inputs will be application specific, so consequently are out of scope for Janus. Rather, Janus in its current form focuses on the agent architecture and its relationship with both the logical architecture and the physical architecture.

## 3. The Agent Architecture

As noted in [6], a cyber-physical agent can be conceptualised as having a behaviour part and an embodiment part. In Janus, this conceptualisation is applied at the architectural level with the agent architecture being mapped to both the physical architecture (embodiment) and the logical architecture (behaviour). In terms of the physical architecture/agent architecture mapping, in Janus, the initial mapping is 1-1, with systems/sub-systems mapping to teams of agents and devices/machines mapping to agents. Behaviours/requirements in the logical architecture are then allocated to teams/agents in the agent architecture. However, in order for a cyber-physical element to exhibit autonomy and thereby provide the flexibility that autonomy entails, Janus defines behaviour in terms of goals, rather than functions. Nonetheless, system goals can be decomposed in a manner analogous to functional decomposition. While goals and functions may appear to be similar, as evidenced by the goal decompositions presented in [2], a key point of differentiation is that goals embody choice – when a goal achievement is delegated to an entity (either an agent or a team of agents), the entity chooses the way in which the goal is achieved depending on the current situational context.

With Janus, the analogy with the traditional functional analysis and allocation process, which is central to systems engineering design, extends beyond decomposition. A key tool in functional analysis is the Functional Flow Block Diagram (FFBD) which details the sequencing of actions (functions) [16]. In an FFBD, functions are represented as behaviour nodes called blocks. Control nodes are then added to sequence the behaviours. In Systems Engineering, FFBDs are used primarily as an aid in function decomposition. However, in Janus, process models (the goal-based analogue of FFBDs) both control nodes and behaviour nodes are modelled as instances of the GORITE framework's Goal class and its sub-classes [2]. Goal-specific behaviour is provided by overriding the Goal.execute() method. Default behaviours are provided for all GORITE control goal classes, which include classes to model choice, sequence, loop and parallel nodes. The resulting process models are executable and BDI execution semantics are preserved. In this regard, if there are multiple plans (models) that could be employed for the achievement of a particular goal (the applicable set), if a plan fails, then the applicable set can be regenerated and (possibly) another plan chosen.

If a leaf node of an agent's process model results in physical action, interfacing to the corresponding device/machine in the physical architecture will be required. In manufacturing, this will typically involve interaction with PLC-controlled devices. As befitting a top-down design methodology grounded in a Systems Engineering approach, Janus is neutral with respect to the actual control technology (e.g. IEC 61131 or IEC 61499) employed.

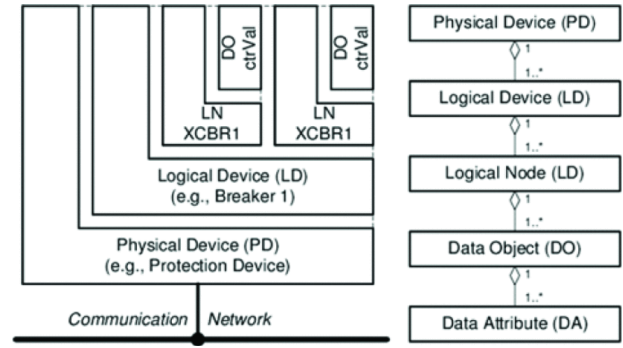

**Figure 1.** IEC 61850 IED Model [17]

When the design of the basic cyber-physical system is complete, goals will have been identified for the primary system function (e.g. product generation / service provision) and sub-goals allocated to each team/agent in the agent architecture. For the leaf nodes, their process models will be grounded in physical action. However, for intermediate elements (i.e. teams) some or all of its elements may not be grounded – a leaf node in a process model for a team can be a goal that is expected to be achieved by another agent or team. Consequently, there is the opportunity for flexibility in terms of how an intermediate process model is achieved, as the goals that it requires to be achieved by other agents are not bound to actual agents in the process model specification. Rather, binding (called task team formation in Janus) is a separate process that can occur prior to or during process model execution. As demonstrated in [2], a task team can be automatically reformed if a current task team member fails and there is a suitable alternative available. Or in the case of [9],[10], if one of the manipulator cylinders fails, then operation can continue, but with a reduced task team size and functionality. Support for both team formation and reformation are provided in the GORITE framework classes.

Process model node behaviours are detailed using the GORITE framework classes and Java code, as described in [2]. In this regard, note that process model execution in GORITE is delegated by the owning team to an executor object. This object then traverses the process model graph. At each node, it invokes the execute() method of the bound goal. Execution is time sliced, so multiple goals can be progressed concurrently. Also, when a node is executed, an object containing data relating to the process model execution is made available to the bound goal, thus providing a business process modelling metaphor for goal achievement. Such a metaphor is particularly apt for product manufacture, where the order being filled can be represented in the data context.

## 4. Power Substation Automation

As indicated in section I, two common types of industrial enterprises are those that produce products and those that provide services. Janus was conceived through consideration of enterprises that produce manufactured products. In this section, we examine its applicability to industrial service provision, in particular, power distribution at the substation level.

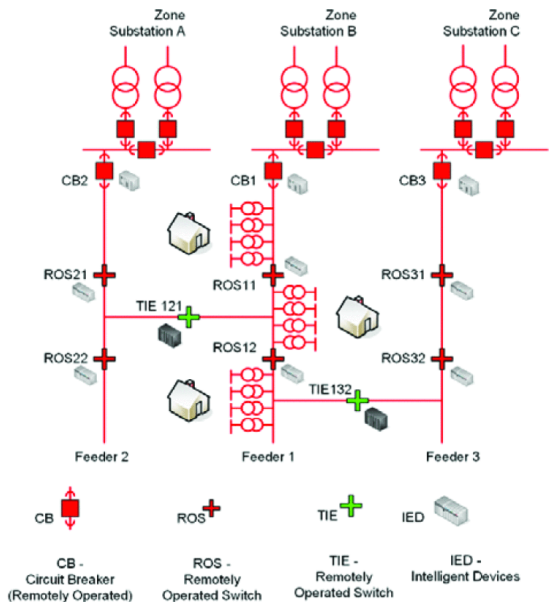

Figure 2. A small distribution utility [18]

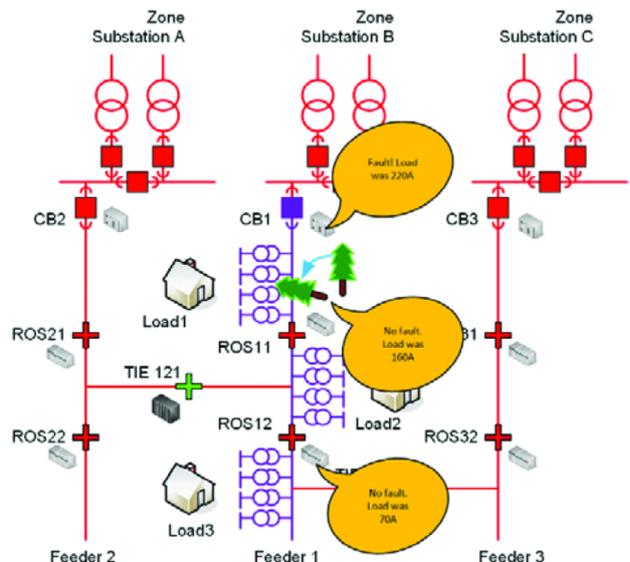

Figure 3. The status of fault detection from CB1, ROS11 and ROS12 in the presence of a permanent fault at location Load1

As mentioned in section 3, Janus is characterised by a top-down systems engineering approach where the overall system specifications and their interactions are defined first (irrespective of the level of decentralization of the system) before working "downwards" in the detailed design of the sub-systems. This approach is in contrast to the current method of engineering substation systems where the design approach is very much bottom-up. This is due to the vendor-driven nature of the industry where the Protection and Control (PAC) functionalities are tightly coupled to the availability of the hardware devices. Ideally, the design of the substation automation system should be driven by the technical requirements for the system based on the need for the protection and the control philosophy rather than the limitations of the chosen or the installed equipment.

The IEC community have made significant advances in addressing the challenges of system design with the introduction of the IEC 61850 standard [17]. This standard defines communication protocols for use by intelligent devices (IEDs) in power substations, together with an abstract data model so that conformant devices are interoperable. Of particular interest from a Janus perspective is that the data model groups standardised data and services into what are called logical nodes (LNs). The LN specification defines semantics of real-time and nonreal-time data which are exchanged between IEC 61850 devices (i.e., IEDs). The different elements of the data model are hierarchically organized as illustrated in Fig. 1. The implementation of LN functionality is not specified by IEC 61850 and various technologies can be employed, such as IEC 61499, as in [18].

In Janus, there are five key steps to the design process:

1. Goal (requirement) decomposition
2. Physical architecture specification
3. Agent team architecture specification
4. Mapping of the agent team architecture to the physical architecture
5. Allocation of goals to agents and teams

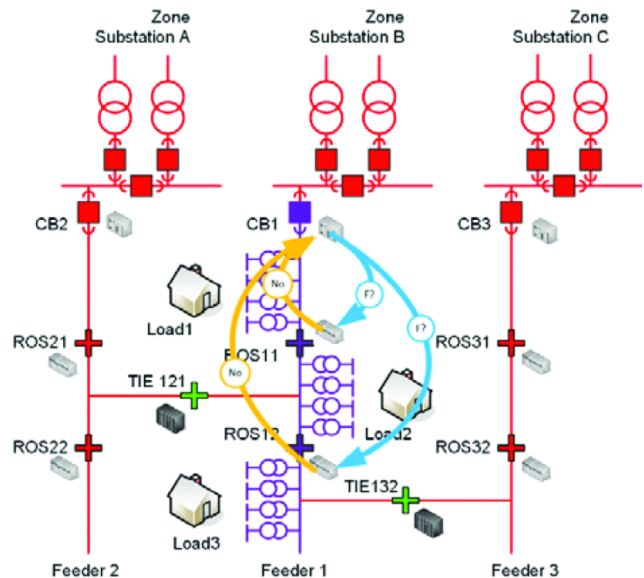

Figure 4. Interactions between the CB1, ROS11 and ROS12 intelligence on locating and isolating the permanent fault

If a substation is to be designed to be IEC 61850 conformant, then the standard provides elements (logical nodes and logical devices) that can be used as leaf nodes in both goal decompositions and physical architecture decompositions. Furthermore, an IEC 61850 logical device can be viewed as a cyber-physical entity which accesses its physical embodiment via logical nodes and interoperates with other cyber-physical entities via its containing physical device/IED. However, the modelling of non-leaf nodes in both the physical and logical (goal) architectures is outside the scope of IEC 61850. With Janus, both physical and logical architecture design follow normal systems engineering practice using a top-down approach and hierarchies will be created that are grounded in logical devices and logical nodes. Also, the agent architecture mirrors the resulting physical architecture, with leaf nodes (cyber-physical entities) mapped to agents and non-leaf nodes (such as feeders) to teams.

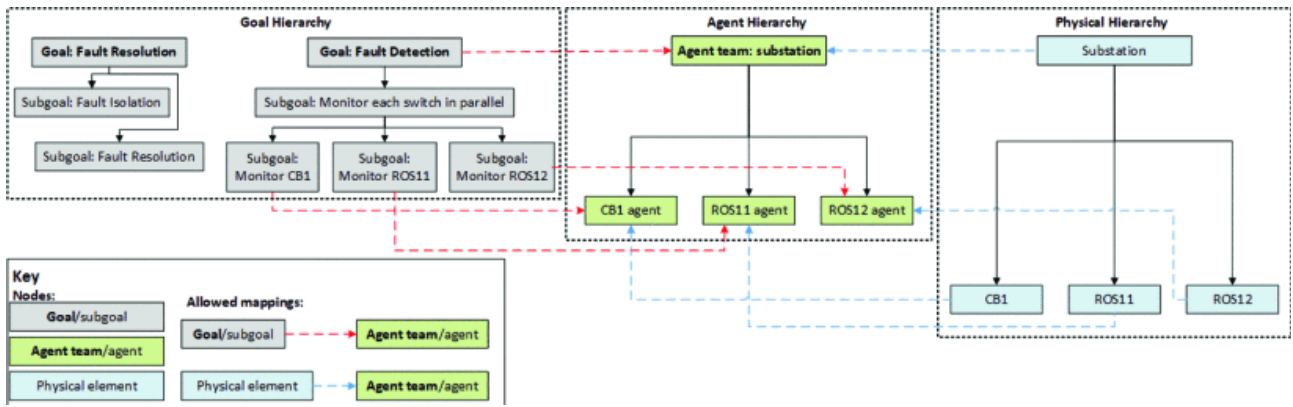

**Figure 5.** Partial mapping of Fault location and isolation agents with respect to physical architecture

Note that a leaf node (i.e. a cyber-physical device) will have autonomy with respect to some functions, such as protection. However, there are other behaviors such as load sharing and Fault Location, Isolation and Service Restoration (FLISR) that involve multiple cyber-physical entities and are therefore allocated to teams.

We will now consider how FLISR operates, using the simple distribution utility illustrated in Fig. 2. This utility consists of three 11kV feeders supplied by three different zone substations. The zone substations are isolated from each other by TIE switches which are set in the open position while the ROS switches are in the closed position. This means that each of the substation zones is operating independently of each other and are responsible for serving their own feeder loads. Each of the switching units is a cyber-physical entity in the sense described previously.

Consider the following scenario. A permanent fault occurs on Feeder 1 of Zone substation B at the location of Load 1 as shown in Fig. 3. This induces a fault current which is detected by the cyber-physical entity CB1. Cyber-physical entities ROS11 and ROS12 do not detect a fault current since they are downstream from the fault location at Load1. Sensing the presence of a fault current, CB1's circuit breaker is immediately "opened" and the power supply is cut to all loads on Feeder1 (i.e. load2 and load3).

At this stage, the Feeder1 team (consisting of CB1, ROS12 and ROS12) does not yet know the exact location of the fault. Therefore, the team initiates the fault isolation process to isolate the fault from the distribution system. It sends query signals to both ROS11 and ROS12 checking to see whether a fault current was detected as shown in Fig. 4. Based on the reply from ROS11 and ROS12, the team can determine the location of the fault. In this scenario, since ROS11 and ROS12 did not sense a presence of a fault current, the Feeder1 team can determine that the fault location is likely to be between CB1 and ROS11. The team will then ask ROS11 to open its switch to isolate the fault from the rest of the distribution system.

The last step in the FLISR scheme is service restoration. Once the fault location is isolated, alternative supply is needed to re-energize Load2 and Load3 respectively. This requires a team operating at a higher level than the feeder teams (e.g. a substation team), as it needs to maintain internal maps of self-restoration routes. Request message will then be propagated through the switching units along these pre-defined restorative paths as shown in Fig. 6. In this scenario, the alternative supply for Load 3 is Zone Substation 3 along the restorative path TIE132, ROS32, ROS31, CB3. After the restoration request is granted by Zone Substation 3, the TIE 132 switch will be closed, transferring Load 3 to Zone Substation 3. To restore service to Load2, the restoration request is sent via the path TIE121, ROS21, CB2.

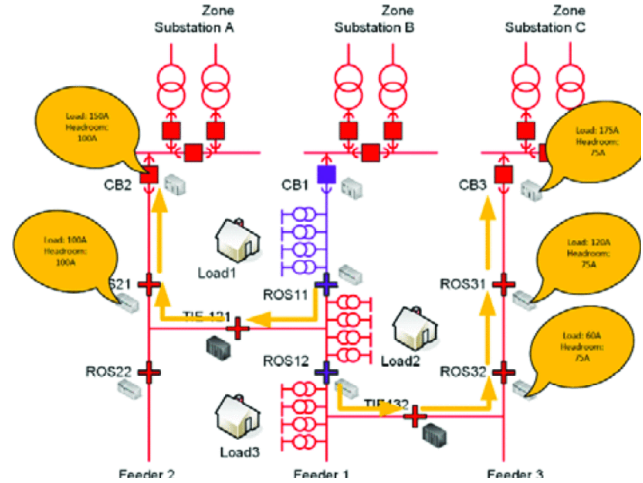

**Figure 6.** Service restoration agent interactions between the switching units

To summarize, the FLISR goal can be decomposed as follows:

*FLISR*

   *fault detection*

      *monitor each switch in parallel*

   *fault resolution*

      *fault isolation*

      *fault restoration*

The FLISR goal and the fault detection, fault resolution and fault restoration goals are allocated to the substation team. Fault isolation goals are allocated to the feeder teams and monitor goals are allocated to switch agents. A partial mapping (focusing on fault detection) is presented in Fig. 5. In the diagram, the physical architecture is collapsed into two levels and the fault resolution goal is not refined or mapped.

Note that the FLISR goal "decomposition" is, in fact, a process model, as it incorporates both implicit (sequential) and explicit (parallel) execution. There is also an understanding that sub-goals will be further refined until actionable behaviors are reached. For example, the fault detection goal could be realized using the following logical nodes:

- TCTR (Non-functional): The current transformer LN is used to measure the current at a fixed sampling rate.
- PIOC (Functional): The overcurrent LN is used to compare the current measurement against a predefined threshold value. If the measured reading exceeds the threshold value, an overcurrent fault is registered.
- PTRC (Functional): The trip conditioning LN is used to check whether the conditions for the tripping of the circuit breaker/switches are satisfied.
- CSWI (Functional): The switch control LN which issues the tripping command to the circuit breaker
- XCBR (Non-functional): The circuit breaker LN issues the physical tripping signal to the Circuit Breakers.

As noted earlier, IEC 61850 does not specify how logical nodes are to be implemented. In this discussion, we have considered logical nodes as providing a functional API for cyber-physical entities with the expectation that a technology appropriate for real-time systems, such as IEC 61499, would be employed. However, there is also coordination of logical node actions to be considered. Whether this coordination is realised by a GORITE process model or through the use of technology such as IEC 61499 will be dependent on performance requirements.

## 5. Discussion

The substation automation case study has demonstrated that a key substation automation task (FLISR) is amenable to design using Janus and IEC 61850. As discussed in section 4, IEC 61850 provides a detailed data model and protocols that enable IEC 61850 compliant devices (which we refer to as cyber-physical entities) to inter-operate. However, the standard does not extend to model and service implementation and importantly is agnostic with respect to implementation technology. Consequently, IEC 61850 integrates seamlessly with the top down design approach that underpins Janus. In contrast, the manufacturing systems considered in Section 2 were not supported by a similarly rich (and standardised) domain model.

One of the key benefits of Systems Engineering in general and Janus in particular, is that its top-down approach can provide design flexibility, as implementation issues (such as choice of technology) and detailed design issues (such as the modelling of logical nodes) can often be deferred to a subsequent design refinement. For example, in [18], the decision was made to employ a multi-agent architecture in which logical nodes were mapped to agents and implemented as IEC 61499 function blocks. With Janus, it may well be that such a design decision is made. However, given the top-down nature of Janus, the decision does not need to be made early in the design process. Also, by deferring this particular decision, its applicability can be constrained to the subsystems/goals that need the performance and flexibility provided by such an approach. For other sub-systems/goals that do not have such stringent requirements, it may be more appropriate to implement their functionality using high-level languages and agent frameworks such as GORITE.

The ability of GORITE agents to interact effectively with devices controlled by IEC 61499 function blocks has been demonstrated in [9], [10]. As highlighted in the previous paragraph, this now means that a designer has numerous options in determining where the boundary between agents/teams and control actions should be placed. Technical performance measures will clearly drive this process, but the top-down nature of Janus encourages the consideration of design alternatives/technologies at the sub-system level and lower. An interesting additional option would be to implement GORITE itself as a function block. Doing so would enable BDI conformant IEC 61499 agents and teams to be constructed. More importantly, it would enable the seamless integration of GORITE agents on both sides of the technology boundary.

## 6. Conclusion and Future Work

With Janus, the design of agent-based cyber-physical systems is viewed as an extension of the traditional Systems Engineering process. This paper has demonstrated that the adoption of such a design stance is feasible. Furthermore, key benefits arising from the top down approach which is central to Systems Engineering (such as late commitment to technology choices) are preserved.

In terms of future work, Janus needs to be deployed in the development of a range of representative industrial systems in order to quantify the benefits of the approach and to identify general guidelines for agent architecture design. Also, tool support in a similar form to the Prometheus Design Tool is required if Janus is to impact on practical system development.